\documentclass[apj,twocolumn]{emulateapj}
\usepackage{amsmath}
\eqsecnum

\renewcommand\email\texttt

\def\spose#1{\hbox to 0pt{#1\hss}}
\def\lta{\mathrel{\spose{\lower 3pt\hbox{$\sim$}}
    \raise 2.0pt\hbox{$<$}}}
\def\gta{\mathrel{\spose{\lower 3pt\hbox{$\sim$}}
    \raise 2.0pt\hbox{$>$}}}

\begin{document} 

\slugcomment{\sc submitted to \it the Astrophysical Journal}
\shorttitle{\sc The Hercules-Aquila Cloud} 
\shortauthors{\sc V.~Belokurov et al.}

\title{The Hercules-Aquila Cloud}

\author{V.\ Belokurov\altaffilmark{1},
N.\ W.\ Evans\altaffilmark{1},
E.\ F.\ Bell\altaffilmark{2},
M.\ J.\ Irwin\altaffilmark{1},
P.\ C.\ Hewett\altaffilmark{1},
S.\ Koposov\altaffilmark{2},
C.\ M.\ Rockosi\altaffilmark{3},
G.\ Gilmore\altaffilmark{1},
D.\ B.\ Zucker\altaffilmark{1}, 
M.\ Fellhauer\altaffilmark{1},
M.\ I.\ Wilkinson\altaffilmark{1},
D.\ M.\ Bramich\altaffilmark{1},
S.\ Vidrih\altaffilmark{1}, 
H.-W.\ Rix\altaffilmark{2},
T.\ C.\ Beers\altaffilmark{4},
D.\ P.\ Schneider\altaffilmark{5},
J.\ C.\ Barentine\altaffilmark{6},
H.\ Brewington\altaffilmark{6},
J.\ Brinkmann\altaffilmark{6},
M.\ Harvanek\altaffilmark{6},
J.\ Krzesinski\altaffilmark{6},
D.\ Long\altaffilmark{6},
K.\ Pan\altaffilmark{6},
S.\ A.\ Snedden\altaffilmark{6},
O.\ Malanushenko\altaffilmark{6},
V.\ Malanushenko\altaffilmark{6}
}

\altaffiltext{1}{Institute of Astronomy, University of Cambridge,
Madingley Road, Cambridge CB3 0HA, UK;\email{vasily,nwe@ast.cam.ac.uk}}
\altaffiltext{2}{Max Planck Institute for Astronomy, K\"{o}nigstuhl
17, 69117 Heidelberg, Germany}
\altaffiltext{3}{Lick Observatory, University of California, Santa
Cruz, CA 95064.}
\altaffiltext{4}{Department of Physics and Astronomy, CSCE: Center for
the Study of Cosmic Evolution, and JINA: Joint Institute for Nuclear
Astrophysics, Michigan State University, East Lansing, MI 48824}
\altaffiltext{5}{Department of Astronomy and Astrophysics,
Pennsylvania State University, 525 Davey Laboratory, University Park,
PA 16802}
\altaffiltext{6}{Apache Point Observatory, P.O. Box 59, Sunspot, NM 88349}

\begin{abstract}
We present evidence for a substantial overdensity of stars in the
direction of the constellations of Hercules and Aquila. The Cloud is
centered at a Galactic longitude of $\ell \approx 40^\circ$ and
extends above and below the Galactic plane by at least $50^\circ$.
Given its off-centeredness and height, it is unlikely that the
Hercules-Aquila Cloud is related to the bulge or thick disk.  More
likely, this is a new structural component of the Galaxy that passes
through the disk.  The Cloud stretches $\sim 80^\circ$ in
longitude. Its heliocentric distance lies between 10 and 20 kpc so
that the extent of the Cloud in projection is $\sim 20$ kpc by $\sim
15$ kpc.  It has an absolute magnitude of $M_v = -13$ and its stellar
population appears to be comparable to, but somewhat more metal-rich
than, M92.
\end{abstract}

\keywords{galaxies: kinematics and dynamics --- galaxies: structure
--- Local Group --- Milky Way:halo}

\section{Introduction}

The construction of the Milky Way via the hierarchical aggregation of
small galaxies has left behind debris in the form of streams,
satellites and substructure in the Galactic halo. There are many
examples now known of such detritus, including the the disrupting
Sagittarius galaxy and its stream~\citep[e.g.,][]{Ib01,Ma03,Be06b},
the ``Monoceros Ring''~\citep[e.g.,][]{Ya03} and the Orphan
Stream~\citep[e.g.,][]{Be07}.  In addition to the easily recognizable
streams, the depth and wide area coverage of the Sloan Digital Sky
Survey (SDSS) enables identification of other less well-defined
features.  A substantial overdensity in the direction of
Virgo~\citep{Ju06} and a possible underdensity in the direction of
Ursa Major~\citep{Xu06} have been found. Here, we identify and
characterize another overdensity of stars, called the {\it Hercules-Aquila
Cloud}.

\section{The Hercules-Aquila Cloud}

\begin{figure*}[t]
\begin{center}
\includegraphics[height=8.5cm]{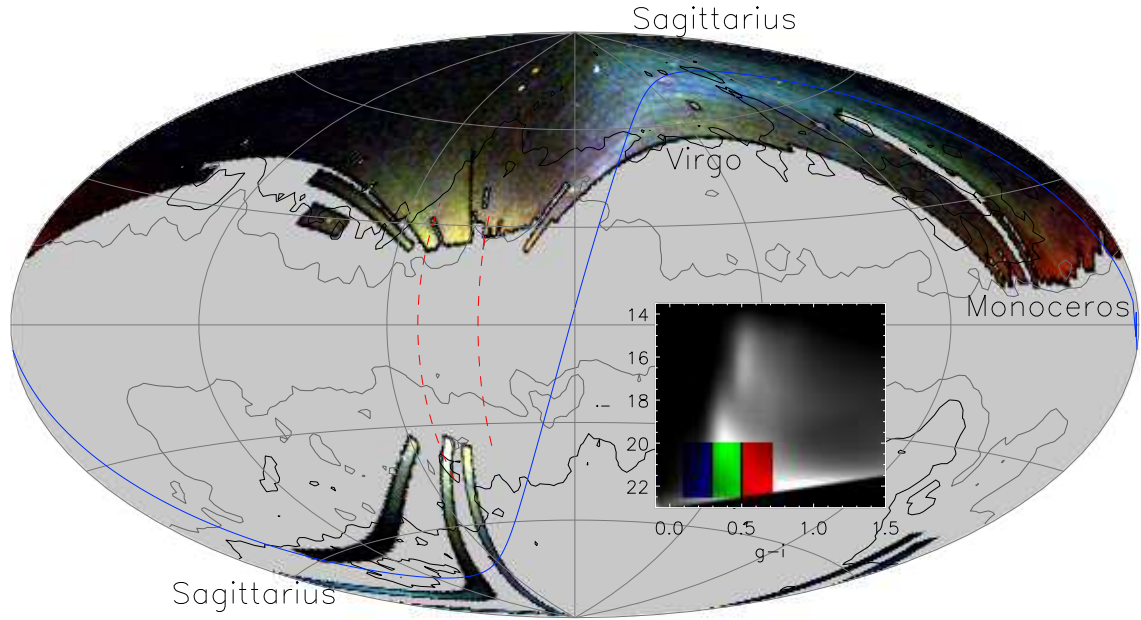}
\caption{\label{fig:bigmap} The areal density of SDSS stars with $0.1
< g-i < 0.7$ and $20 < i < 22.5$ in Galactic coordinates. The color
plot is an RGB composite with blue representing $0.1 \le g-i < 0.3$,
green $0.3 \le g-i <0.5$ and red $0.5\le g-i < 0.7$. Structures
visible on the figure include the arch of the Sagittarius stream and
the Orphan Stream in blue and the multiple wraps of the Monoceros ring
in red. Behind, and offset from, the Galactic bulge, there is a
yellow-colored structure visible in SDSS data in both hemispheres,
centered at $\ell \approx 40^\circ$. This is the Hercules-Aquila
Cloud.  Overlaid on the plot are contours of constant extinction
(black is 0.1 mag and grey is 0.25 mag in the $i$ band). The two red
vertical lines show the SEGUE stripes at $\ell = 31^\circ$ (runs 5378
and 5421) and $50^\circ$ (run 4828). The inset is a color-magnitude
diagram of the SDSS data, with blue, green and red bands encompass the
stars used to create the large figure.}
\end{center}
\end{figure*}
\begin{figure*}
\begin{center}
\includegraphics[height=8.5cm]{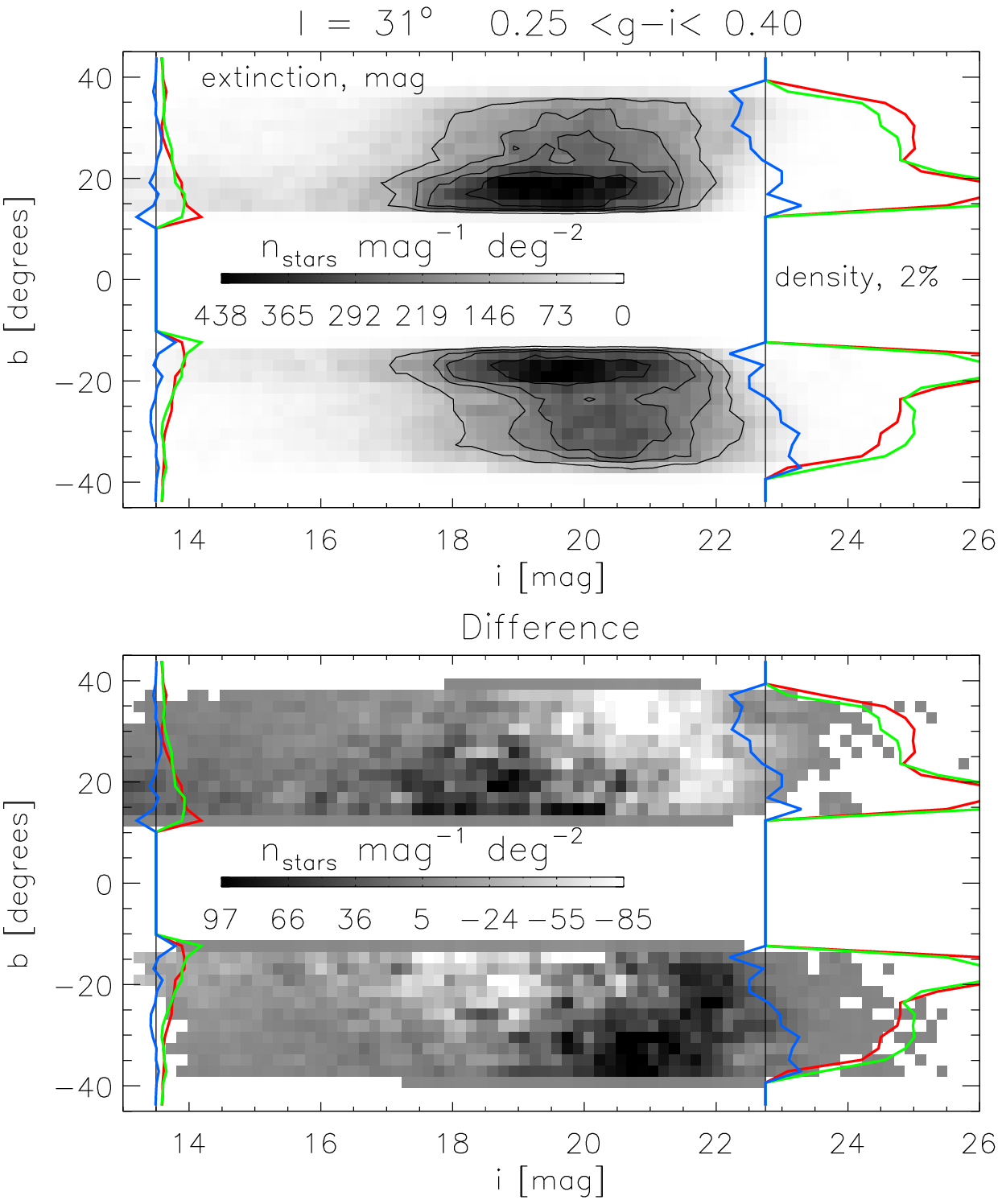}
\includegraphics[height=8.5cm]{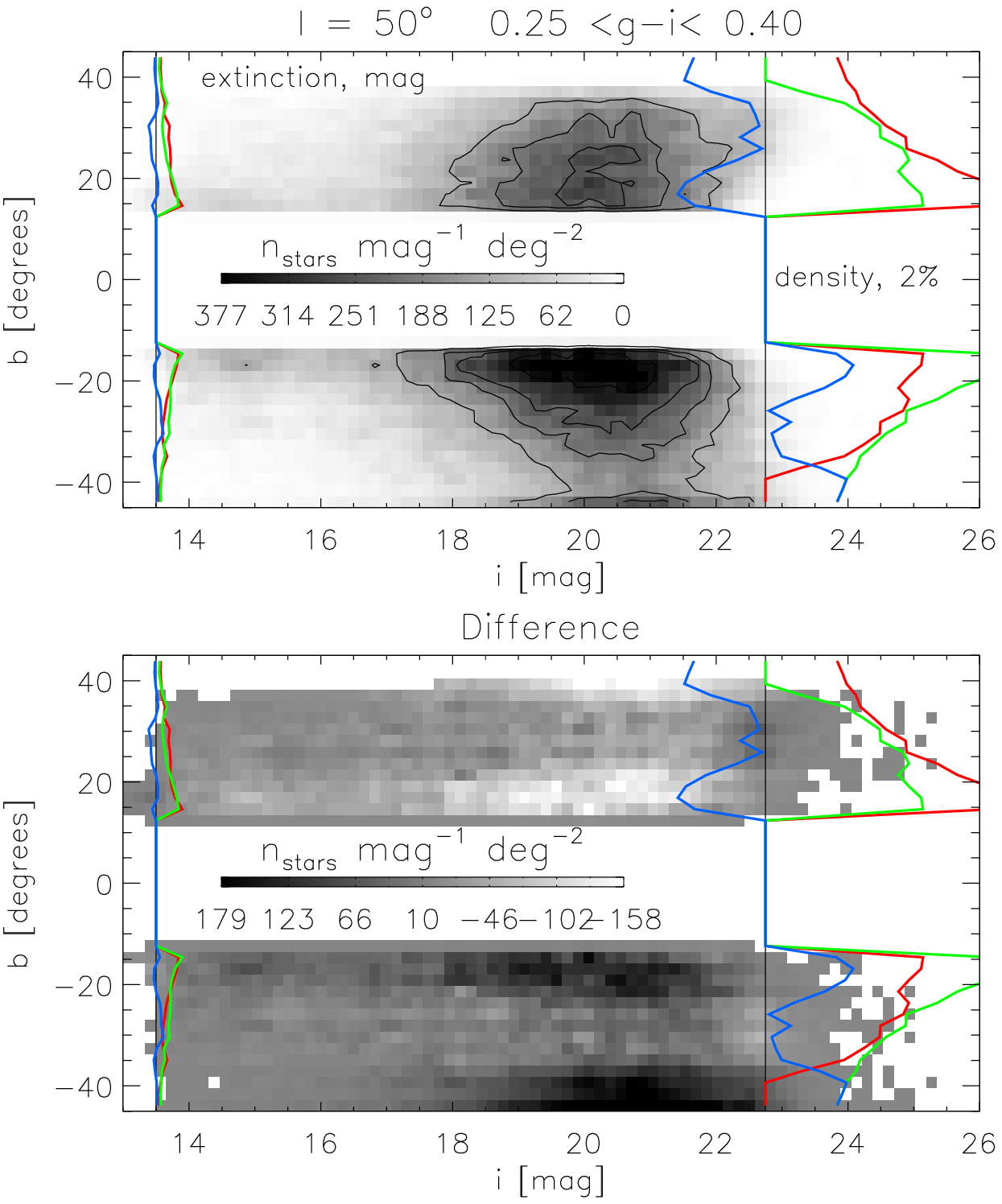}
\caption{\label{fig:slicea} The upper panels show greyscale contours
of the spatial density of SDSS stars with $0.25 < g -i < 0.4$ in two
slices at longitude $\ell = 31^\circ$ (left) and $\ell = 50^\circ$
(right). The spatial density is measured in ${\rm mag}^{-1} {\rm
deg}^{-2}$.  In each case, the leftmost (rightmost) colored lines
shows the variation of extinction (density) with latitude in each
hemisphere. Green denotes the current profile, whilst red refers to
the opposite hemisphere and blue the difference. There are a number of
prominent enhancements of density along the line of sight, as
manifested by green-colored profiles, which are not smoothly decaying
curves but show bumps.  The southernmost overdensities are associated
with the Sagittarius stream ($\ell = 31^\circ, b \approx -35^\circ, i
\approx 20.5$ mag and $\ell = 50^\circ, b \approx -40^\circ, i \approx
20.5$ mag). The other enhancement is part of the Hercules-Aquila
Cloud, which is revealed in the difference plots shown in the lower
panels. These are created by subtracting the density in the two
hemispheres so that black (white) indicates relative overdensity
(underdensity). The Hercules-Aquila Cloud is at $\ell = 31^\circ, b
\approx 20^\circ, i \approx 19.$ mag and $\ell = 50^\circ, b \approx
-15^\circ, i \approx 20.$ mag.}
\end{center}
\end{figure*}
\begin{figure}[t]
\begin{center}
\includegraphics[width=8.cm]{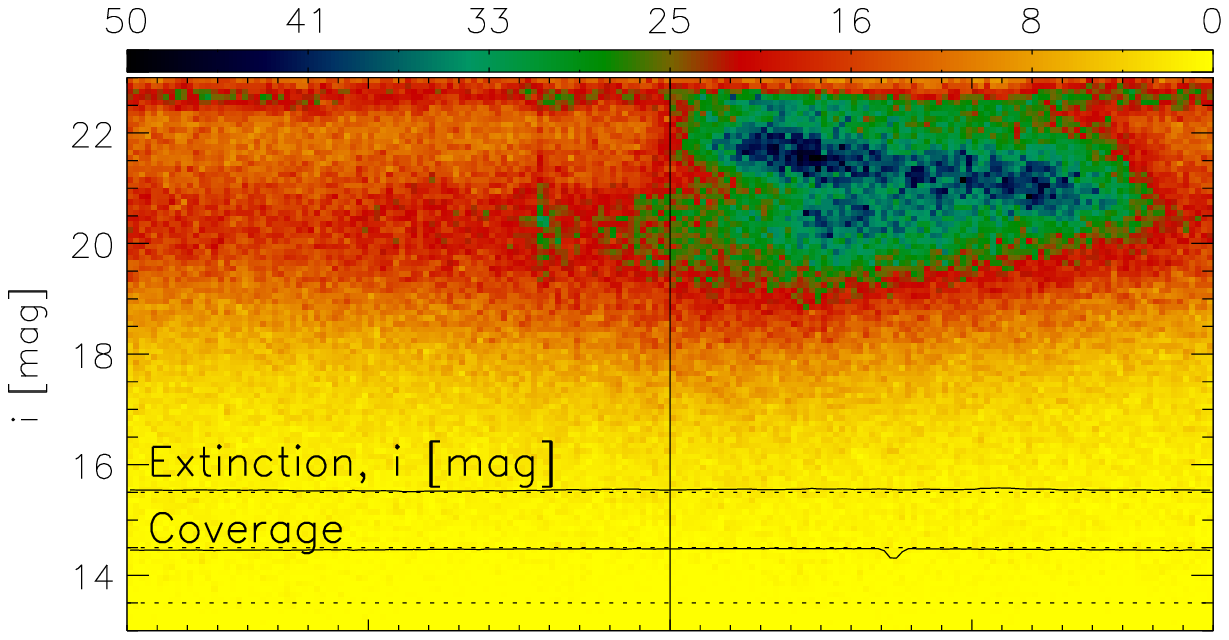}
\includegraphics[width=8.cm]{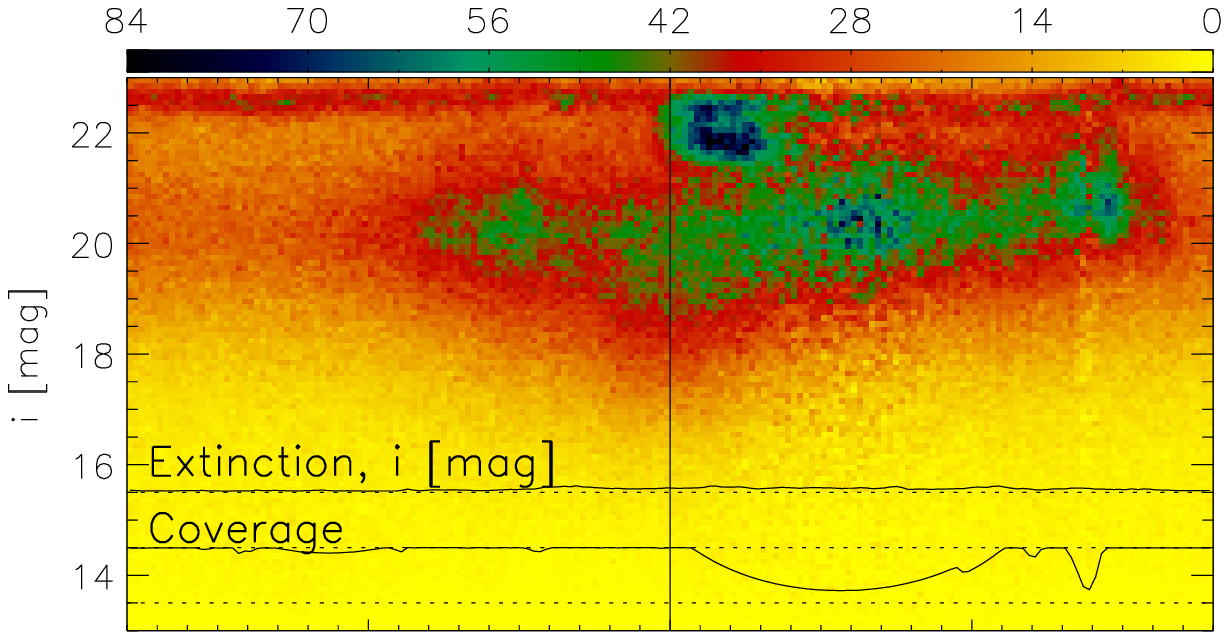}
\includegraphics[width=8.cm]{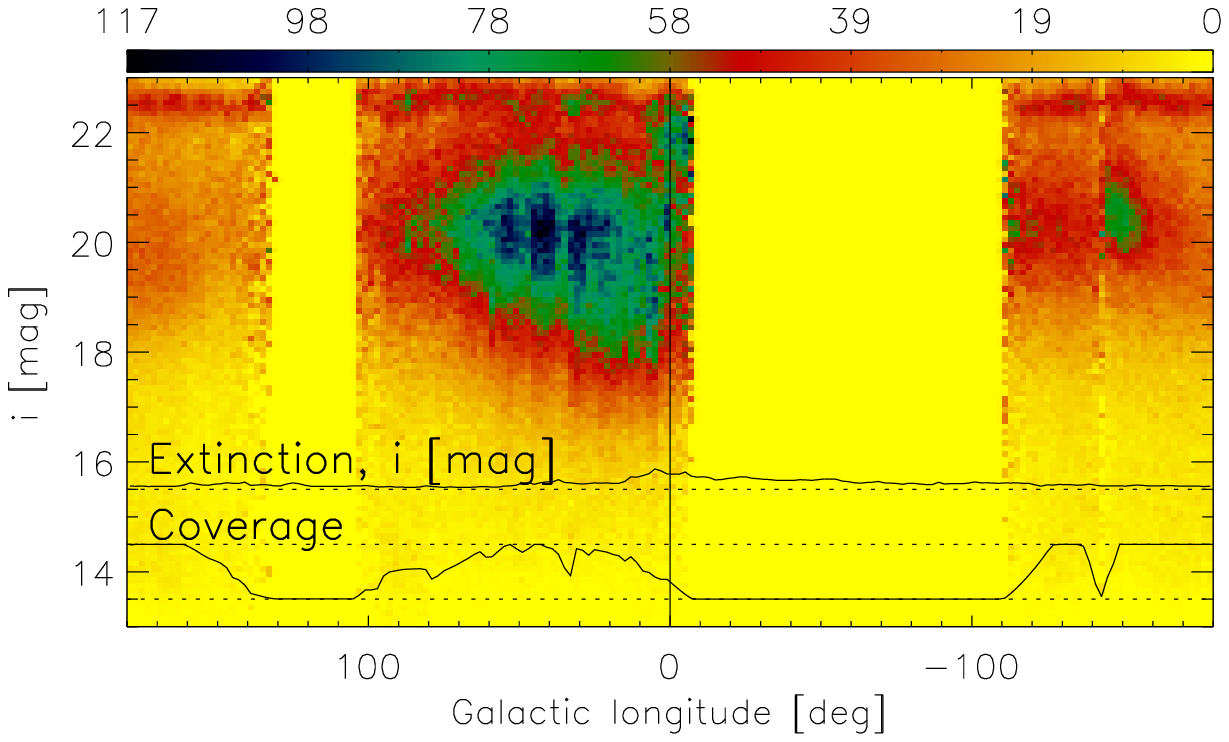}
\caption{\label{fig:lonmap} Three density slices at $b = 70^\circ$
(top), $55^\circ$ (middle) and $40^\circ$ (bottom), each of which is
$15^\circ$ wide. The key gives the scale in number of stars ${\rm
mag}^{-1} {\rm deg}^{-2}$ satisfying the color cut $0.25 <g-i
<0.4$. The variation of the extinction along the strips and the
fraction (between 0 and 1) covered in DR5 are shown as a function of
longitude at the bottom of the panels. Three structures are
discernible in the panels and they are clearly separated. In the top
panel, the Sagittarius stream and Virgo Overdensity are seen to cross
each other obliquely, whilst the tip of the Hercules-Aquila Cloud is
just visible; in the middle panel, the Sagittarius stream is
bifurcated and cleanly differentiated from the Virgo Overdensity,
whilst the Hercules-Aquila Cloud is becoming more substantial; in the
lowest panel only the Hercules-Aquila Cloud is seen, due to SDSS DR5
coverage.}
\end{center}
\end{figure}
\begin{figure}[t]
\begin{center}
\includegraphics[width=9.cm]{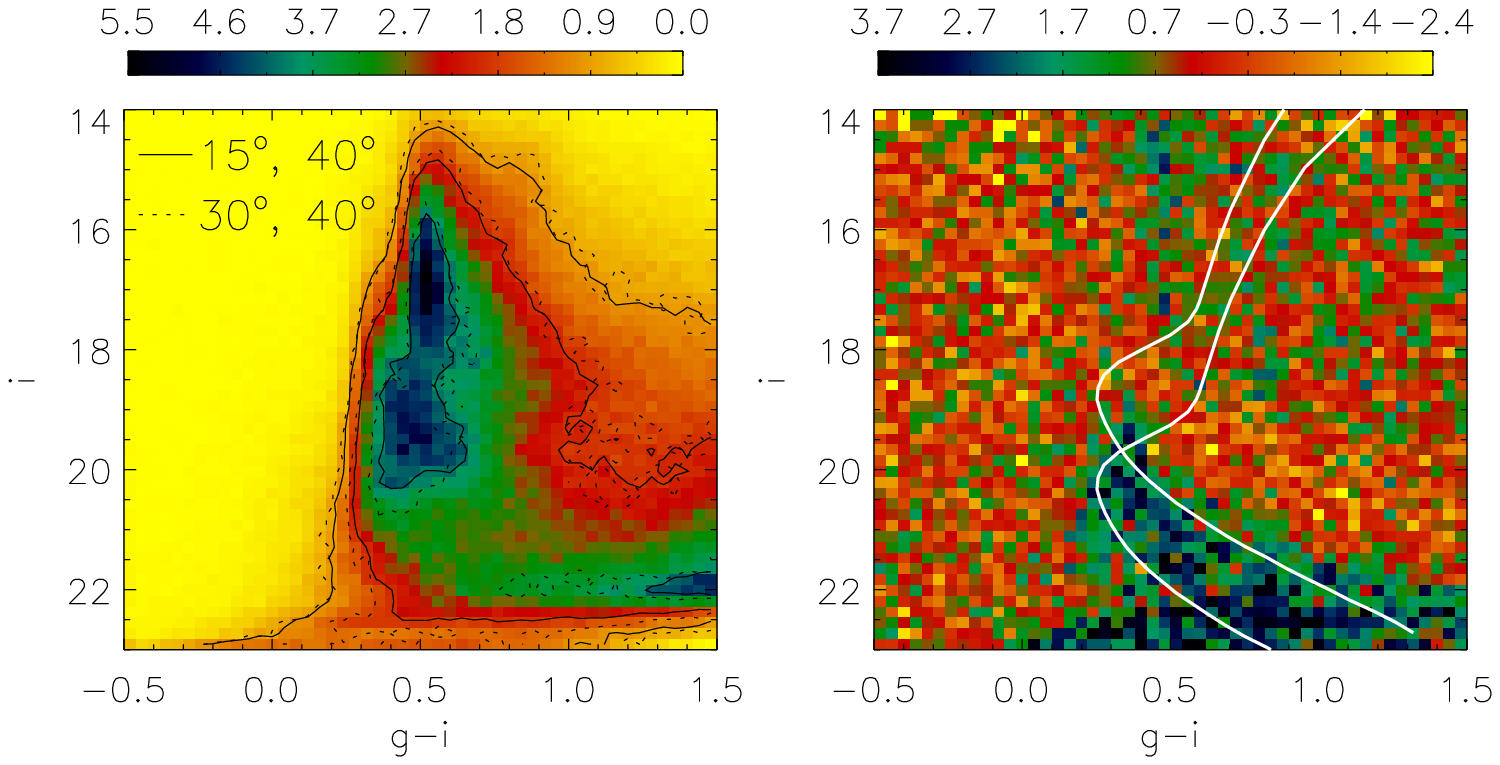}
\caption{\label{fig:cmd} Left: Hess diagrams for the on-Cloud field at
$\ell = 30^\circ, b=40^\circ$ (dotted lines), and the off-Cloud field
at $\ell = 15^\circ, b=40^\circ$ (solid).  The units are $10^4$ stars
per square magnitude.  Right: Difference of the Hess diagrams divided
by the square root of their sum.  This shows essentially the
signal-to-noise ratio in the difference. There is one obvious
overdensity corresponding to upper main sequence and turn-off stars in
the Hercules-Aquila Cloud. The white lines show M92 ridgelines shifted
to the distance of 10 and 20 kpc, which bracket the main sequence
location.}
\end{center}
\end{figure}

SDSS imaging data are produced in five photometric bands, namely $u$,
$g$, $r$, $i$, and $z$~\citep[see e.g.,][]{Fu96,Ho01,Sm02,Gu06} and
are automatically processed through pipelines to measure photometric
and astrometric properties and to select targets for spectroscopic
follow-up~\citep{Lu99,St02,Pi03,Iv04}.  In this paper, all the
photometric data are de-reddened using the maps of~\citet*{Sc98}.  We
use primarily Data Release 5 (DR5), which covers $\sim 8000$ square
degrees around the Galactic North Pole, and 3 stripes in the Galactic
southern hemisphere~\citep{AM07}.  Whilst the motivation for the
original SDSS survey was largely extragalactic, a substantial portion
of its successor survey SDSS-II is devoted to Galactic structure (the
Sloan Extension for Galactic Understanding and Exploration or
SEGUE). There are a number of SEGUE imaging stripes that run from
positive to negative latitudes through the disk at constant longitude.

Fig.~\ref{fig:bigmap} shows an RGB composite density map of stars with
$20 < i < 22.5$ and $0.1 < g-i < 0.7$ in Galactic coordinates centered
at $(\ell =0^\circ, b= 0^\circ$).  As illustrated in the inset, the
blue color represents the density of stars with $0.1 \le g-i < 0.3$,
green those with $0.3 \le g-i <0.5$ and red those with $0.5\le g-i <
0.7$. A number of the prominent substructures that dominate the halo
are marked, including the Sagittarius Stream, the Monoceros Ring, the
Virgo Overdensity and the Orphan Stream.  The subject of this {\it
Letter} is the gold-colored overdensity that is most prominent in DR5
at longitudes of $\sim 40^\circ$ and is visible in both hemispheres.
This is the Hercules-Aquila Cloud.  Overlaid on the plot are contours
of constant total extinction from~\citet{Sc98}.

To investigate this structure, we exploit the SEGUE imaging at $\ell =
31^\circ$ (runs 5378 and 5421) and $50^\circ$ (run 4828), shown as
the two red dashed lines in Fig.~\ref{fig:bigmap}. Each SDSS and
SDSS-II stripe comprises two interlocking runs, each of which have a
width of $2.34^\circ$ but covers half the area of the stripe.  In
Fig.~\ref{fig:slicea}, we show the density of stars in the runs with
$0.25 < g-i < 0.4$ in units of ${\rm mag}^{-1}{\rm deg}^{-2}$. The
color cut is designed to identify primarily halo stars. We also show
profiles of density and extinction as well as their differences
(in red, green and blue lines respectively). We see that, comparing
the Northern and Southern hemispheres, there is dramatic asymmetry at
$\ell = 50^\circ$ with a substantial overdensity at negative
latitudes. The asymmetry is still present at $\ell = 31^\circ$, but
the overdensity now occurs at positive latitudes.

The structure is seen most cleanly in the difference maps in the lower
two panels of Fig.~\ref{fig:slicea}. Here, overdensities relative to
the other hemisphere are dark, while underdensities are white.  The
Hercules-Aquila Cloud is seen clearly at $\ell = 31^\circ, b \approx
20^\circ, i \approx 19$ mag and $\ell = 50^\circ, b \approx
-15^\circ, i \approx 20$ mag. It is therefore closer to the Sun at
$\ell =31^\circ$ than at $\ell = 50^\circ$.  There are two further
overdensities visible in Fig~\ref{fig:slicea} at $\ell = 31^\circ, b
\approx -35^\circ, i \approx 20.5$ mag and $\ell = 50^\circ, b \approx
-40^\circ, i \approx 20.5$ mag. They are most likely associated with
the Sagittarius stream.  The orbital plane of the Sagittarius dSph as
computed in \citet{Fe06} is shown as a blue dotted line in
Fig~\ref{fig:bigmap}. Its tidal debris has a thickness of $\sim
15^\circ$ on the sky and so should intersect the SEGUE imaging runs at
their southernmost latitudinal points.

In the slice at $\ell = 50^\circ$, the total number of excess stars
with the color cut $0.25 < g-i < 0.4$, the latitude cut $-35^\circ <
b<0^\circ$, and within the contour corresponding to $n_{\rm stars} =
50$ in the lower right panel of Fig~\ref{fig:slicea} is $10^4$. This
ignores the contribution above the plane, and so is a conservative
estimate.  Scaling to the overall longitudinal extent of the Cloud, we
estimate that there are $5 \times 10^5$ stars with $0.25 < g-i <
0.4$. Comparing with the CMD of the metal-poor globular cluster M92,
we etsimate an absolute magnitude of $M_v = -13$ for the overdensity

Fig~\ref{fig:lonmap} shows longitudinal cuts through the
Hercules-Aquila Cloud at three latitudes $b= 70^\circ, 55^\circ$ and
$40^\circ$, each $15^\circ$ wide.  The SDSS data sometimes only partly
cover the slice. To take this incompleteness into account, we
normalize the density in each longitude bin by the fraction covered.
There are three structures visible in the panels.  The bifurcated
Sagittarius stream and the Virgo Overdensity dominate the top two
panels. However, in both panels, a new structure is becoming visible;
this is the Hercules-Aquila Cloud.  In the bottom panel, the Cloud has
come clearly into view, extending over $90^\circ$ in longitude, whilst
the Sagittarius and the Virgo Overdensity are largely missing in the
area covered by the DR5 footprint.

Fig.~\ref{fig:cmd} shows Hess diagrams for a $8^\circ \times 8^\circ$
field centered on $\ell = 30^\circ, b=40^\circ$, and a $16^\circ
\times 16^\circ$ field centered on $\ell = 15^\circ, b=40^\circ$
(left), together with their difference divided by the square root of
their sum (right). The two fields are chosen to lie interior to and
exterior to the Cloud at the same latitude. In the difference panel,
there is a clear upper main sequence and turn-off, corresponding to
stars in the Cloud.  Also shown in white are two M92 ridgelines, using
data from \citet{Cl05}, that are used to bracket the distance range in
the Hercules-Aquila Cloud. These suggest that the Cloud in SDSS DR5 is
primarily composed of upper main sequence and turn-off stars at
heliocentric distances of between 10 and 20 kpc. The turn-off lies
redward of the M92 ridgeline, suggesting that the Cloud is composed of
more metal-rich stars than M92 ([Fe/H] = -2.2).

\begin{figure}[t]
\begin{center}
\includegraphics[width=7cm]{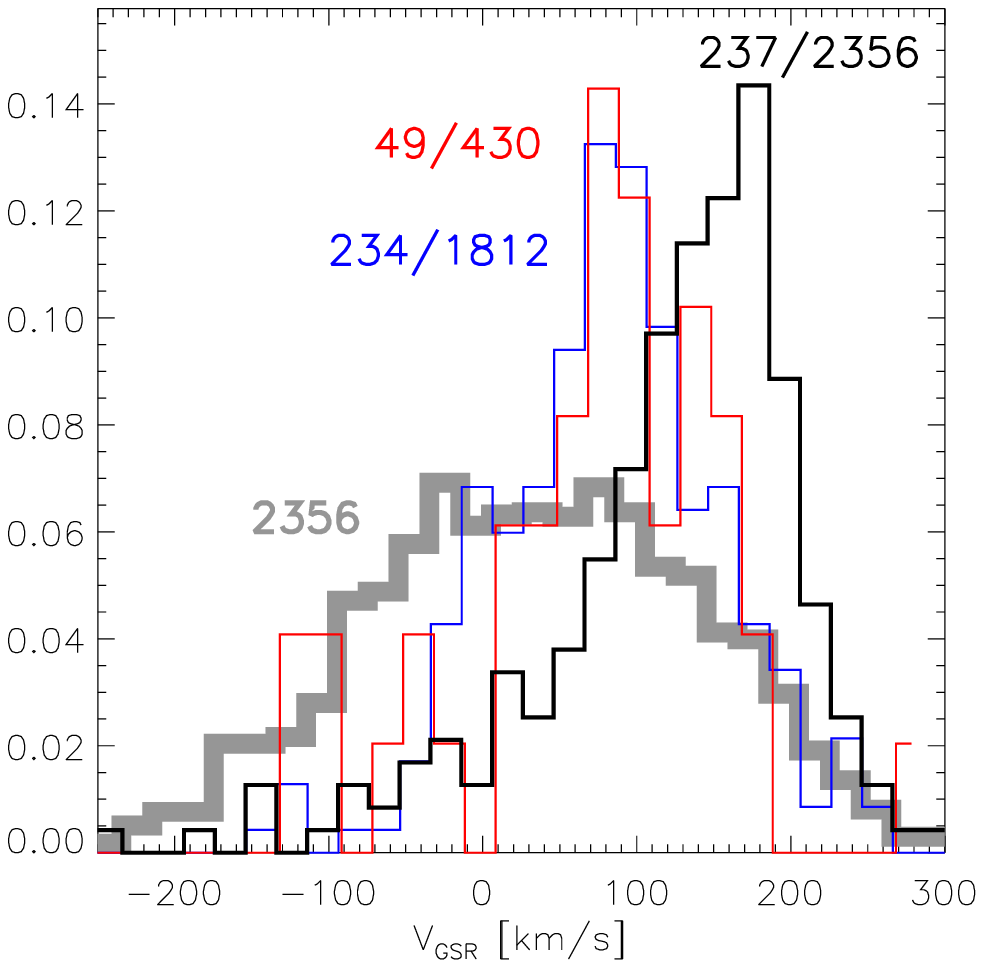}
\caption{\label{fig:vel} Histograms of velocities (converted to the
Galactic standard of rest) for stars with $g-r <1$ and $r < 19$. Thick
grey lines refers to all stars in the field centered on the
Aquila-Hercules Cloud. We select red giant branch candidates from the
CMD in three fields, two off-Cloud (red and blue) and one on-Cloud
(black). The red and blue histograms peak at $\sim 80$ kms$^{-1}$, as
expected for thick disk stars. The black histogram peaks at $\sim 180$
kms$^{-1}$, which we interpret as the velocity signature of the
Cloud. The total numbers of stars in the fields after the CMD cut, as
well as those picked up by the CMD mask, are marked in color above the
histograms.}
\end{center}
\end{figure}

Fig.~\ref{fig:vel} shows the kinematic evidence for the
Hercules-Aquila Cloud using SDSS radial velocities for stars with $g-r
< 1$ and $r < 19$. All the velocities are converted to the Galactic
standard of rest and have a typical uncertainty of $\sim 10$
kms$^{-1}$. The first field covers the Cloud ($20^\circ \le b \le
55^\circ$ and $20^\circ \le \ell \le 75^\circ$). There are two
comparison fields at larger longitude and higher latitude, namely
field 2 at $40^\circ \le b \le 55^\circ$ and $110^\circ \le \ell \le
180^\circ$, and field 3 at $65^\circ \le b \le 80^\circ$ and $20^\circ
\le \ell \le 80^\circ$.  Most of the stars are at high latitude and
selected to be bluer than the thin disk dwarfs. The signal is
therefore dominated by halo and thick disk stars. This is clear from
the thick grey histogram in Fig.~\ref{fig:vel}, which reports
velocities for all stars in Field 1. The distribution is built from
one component with almost zero mean velocity and another with mean at
$\sim 80$ kms$^{-1}$. The all-star velocity distribution is the same
for Fields 2 and 3, and not shown. To pick out the stars in the
structure, we use a mask based on the M92 ridgeline, shifted to 10 and
20 kpc. The distribution for field 1 centered on the Hercules-Aquila
Cloud is shown as a black solid line and can be compared to the
histograms for the off-Cloud fields 2 and 3 (blue and red
respectively). Nearby thick disk stars in these locations have a line
of sight velocity of $\sim 150 \sin \ell \cos b \sim 80$ kms$^{-1}$
with respect to the Galactic standard of rest~\citep[e.g.,][]{Gi89},
which is consistent with the peaks in the red and blue
histograms. There is a clear kinematic signal from stars moving
coherently with a radial velocity of $\sim 180$ kms$^{-1}$ in the
on-Cloud field. This cannot be due to either the halo or the thick
disk.

Could the Hercules-Aquila Cloud be related to the Galactic bulge or
bar? There is an asymmetry in the integrated light of the bulge,
famously seen by the DIRBE instrument on the COBE satellite
~\citep{Dw95}, but it is only of the order of a few degrees at
most~\citep[see e.g.,][]{We94}. This asymmetry would be larger if
confined to stars in a range of heliocentric distances between 4 to 8
kpc, as we would then be primarily picking up stars on the near-side
of the bar. However, our estimate of the distance to the
Hercules-Aquila Cloud is 10 to 20 kpc.  At $\ell = 40^\circ$, the
closest the Cloud can be to the Galactic Center is in the range $6$ to
$15$ kpc. Although the Cloud passes through the disk, it seemingly
lies well beyond the Galactic bar. Also, the Cloud extends to Galactic
latitudes of $\pm 50^\circ$, again much larger than the vertical
scale-height of the bulge or of the thin and thick disks.
\citet{Pa03,Pa04} also discovered some asymmetries in the distribution
of bright, probably thick disc, stars using POSS 1 data, but these
stars are much closer at $\sim 2$ kpc from the Sun.

Could the Hercules-Aquila Cloud be a smooth component that is centered
on $\ell=0^\circ$ but made to appear asymmetric by extinction?  Most
of the extinction is caused by dust with a scalelength of $\sim 100$
pc. However, there is a prominent North Polar Spur of dust that
extends to high Galactic latitudes~\citep[see e.g.,][]{Fr81}. This can
be seen in Fig.~\ref{fig:bigmap} as the extension to the black contour
over latitudes $30^\circ$ to $60^\circ$, where it overlaps part of the
Sagittarius stream. However, a similar amount of extinction appears to
have no effect on the underlying density of detected stars in the
Sagittarius stream and the Virgo Overdensity at $\ell = -100^\circ$,
and so variable extinction is not a plausible explanation of the
offset structure. {\it We conclude that the Hercules-Aquila Cloud is a
real, hitherto unknown, structural component of the inner halo of the
Galaxy.}




\acknowledgments Funding for the SDSS and SDSS-II has been provided by
the Alfred P.  Sloan Foundation, the Participating Institutions, the
National Science Foundation, the U.S. Department of Energy, the
National Aeronautics and Space Administration, the Japanese
Monbukagakusho, the Max Planck Society, and the Higher Education
Funding Council for England. The SDSS Web Site is
http://www.sdss.org/.
                                                                              
The SDSS is managed by the Astrophysical Research Consortium for the
Participating Institutions. The Participating Institutions are the
American Museum of Natural History, Astrophysical Institute Potsdam,
University of Basel, Cambridge University, Case Western Reserve
University, University of Chicago, Drexel University, Fermilab, the
Institute for Advanced Study, the Japan Participation Group, Johns
Hopkins University, the Joint Institute for Nuclear Astrophysics, the
Kavli Institute for Particle Astrophysics and Cosmology, the Korean
Scientist Group, the Chinese Academy of Sciences (LAMOST), Los Alamos
National Laboratory, the Max-Planck-Institute for Astronomy (MPIA),
the Max-Planck-Institute for Astrophysics (MPA), New Mexico State
University, Ohio State University, University of Pittsburgh,
University of Portsmouth, Princeton University, the United States
Naval Observatory, and the University of Washington.

\end{document}